\documentclass[useAMS,usenatbib]{mn2e}

\usepackage{amsmath}
\usepackage{amssymb}
\usepackage{soul}
\usepackage{natbib}
\usepackage{graphicx}
\usepackage{caption}
\usepackage{subcaption}
\usepackage{enumerate}
\usepackage{datetime}
\usepackage{color}
\usepackage{hyperref}
\usepackage{draftcopy}

\definecolor{dylan}{rgb}{0.431,0.106,0.537}

\definecolor{revision}{rgb}{0.9,0.0,0.0}

\begin{document}

\title[Ablation in very massive star formation]
{Line-driven ablation of circumstellar discs: IV. The role of disc ablation in massive star formation and its contribution to the stellar upper mass limit}

\author[N. D. Kee and R. Kuiper]
{Nathaniel Dylan Kee\thanks{Email: nathaniel-dylan.kee@uni-tuebingen.de} and Rolf Kuiper\\
 Institut f\"ur Astronomie und Astrophysik, Eberhard Karls Universit\"at T\"ubingen, Auf der Morgenstelle 10, 72076 T\"ubingen, Germany\\
\\ 
 }

\def\Rstar{R_{\ast}}
\def\Mstar{M_{\ast}}
\def\Lstar{L_{\ast}}
\def\Tstar{T_{\ast}}
\def\gstar{g_{\ast}}
\def\vth{v_{th}}
\def\grad{g_{rad}}
\def\glines{g_\mathrm{lines}}
\def\Mdot{\dot M}
\def\mdot{\dot m}
\def\yr{{\rm yr}}
\def\ksec{{\rm ksec}}
\def\kms{{\rm km/s}}
\def\qad{\dot q_{ad}}
\def\qlines{\dot q_\mathrm{lines}}
\def\solar{\odot}
\def\Msun{M_{\solar}}
\def\msbyr{\Msun/\yr}
\def\Rsun{R_{\solar}}
\def\Lsun{L_{\solar}}
\def\Be{{\rm Be}}
\def\Rpole{R_{\ast,p}}
\def\Req{R_{\ast,eq}}
\def\Rmin{R_{\rm min}}
\def\Rmax{R_{\rm max}}
\def\Rstag{R_{\rm stag}}
\def\vinf{V_\infty}
\def\Vrot{V_{rot}}
\def\Vcrit{V_{crit}}
\def\d{\mathrm{d}}
\def\rres{\mathbf{r}_\mathrm{res}}
\def\ro{\mathbf{r}_\mathrm{loc}}
\def\rs{\mathbf{r}_\mathrm{s}}
\def\req{\mathbf{r}_\mathrm{eq}}
\def\o{\mathrm{o}}
\def\d{\mathrm{d}}
\def\e{\times 10^}

\maketitle

\begin{abstract}
Radiative feedback from luminous, massive stars during their formation is a key process in moderating accretion onto the stellar object.
In the prior papers in this series, we showed that one form such feedback takes is UV line-driven disc ablation.
Extending on this study, we now constrain the strength of this effect in the parameter range of star and disc properties appropriate to forming massive stars.
Simulations show that ablation rate depends strongly on stellar parameters, but that this dependence can be parameterized as a nearly constant, fixed enhancement over the wind mass loss rate, allowing us to predict the rate of disc ablation for massive (proto)stars as a function of stellar mass and metallicity.
By comparing this to predicted accretion rates, we conclude that ablation is a strong feedback effect for very massive (proto)stars which should be considered in future studies of massive star formation.
\end{abstract}

\begin{keywords}
circumstellar matter --
accretion, accretion discs --
stars: massive --
stars: winds, outflows --
stars: formation
\end{keywords}
 
\section{Introduction}
\label{sec:intro}

Observations of populations of massive stars firmly suggest that there is a fundamental upper limit to the mass of a star \citep{Fig05}, with this limit perhaps being as high as several hundred solar masses \citep{CroSch10,SchSan18}.
While the origin of this stellar upper mass limit is not entirely clear, three main possibilities are commonly invoked.
The first of these is the impact of the extreme luminosity of the star on the stability of its structure \citep[e.g.][and references therein]{Owo15}.
The other two possibilities are both consequences of the star formation process, specifically that the stellar upper mass limit is a by-product of large scale fragmentation of the cloud out of which the star is forming or that it is caused by radiative feedback of the forming massive star onto this protostellar cloud \citep[e.g.][and references therein]{ZinYor07}.
While it is very plausible that all three effects play a role, for this work we focus on the final hypothesis of radiative feedback during the star formation process.
% of radiative feedback during star formation.

Previous studies of these formation processes for massive stars already extensively examine the role of feedback from the forming stars both in radiation (predominantly radiation pressure and photoionization) and mechanical injection of momentum and energy (predominantly from jets, winds, and supernovae) into the star's natal environment at pc to kpc scales \citep[e.g.][]{WanLi10,GatWal16,VazGon16}, as well as investigating the role such feedback plays in moderating accretion onto the (proto)star\footnote{We adopt the terminology ``(proto)star'' here to acknowledge that the most massive stars will contract to have a main-sequence-like mass-radius relation well before the full stellar mass is assembled, resulting in hydrogen core burning stars still undergoing vigorous accretion \citep{Kah74,HosOmu09,HosYor10}.} from pc to au scales \citep[e.g. K\"olligan \& Kuiper, submitted;][]{KlaPud16,RosKru16,TanTan17,KuiHos18}.
Due to the incompatibility in numerical simulations between small resolution elements and large dynamic ranges in space and time, however, the final accretion from length scales of order au to the surface of the (proto)star is often omitted.

Generally this omission is not considered to be a fundamental issue for the simulations considered, as the majority of the physics considered occurs on larger scales.
Additionally, these small regions are currently impossible to probe observationally.
However, as discussed in the prior papers in this series \citep[][hereafter papers I, II, and III respectively]{KeeOwo16a,KeeOwo18a,KeeOwo18b} this near-star region within only a few stellar radii of the stellar surface is subject to intense UV irradiation, leading to not only the canonically known line-driven hot star winds but also allowing for the ablation of the surface layers off circumstellar discs.
As the next step in this series, we here examine the potential role that this line-driven disc ablation plays in reducing accretion onto the most massive stars during their formation epochs.
To do so, we here take the approach of focusing on only these final miles of accretion.
In section \ref{sec:param_study} we present the results of our parameter study of ablation for a grid of stellar masses and circumstellar disc masses.
Given the uniformity of behavior found in these simulations, we generalize this in section \ref{sec:upper_mass_lim} to an analytic prediction of disc ablation rate as a function of stellar parameters and as a function of metallicity.
In this section, we also examine the role that disc ablation may play in helping to set the stellar upper mass limit.
Finally, in section \ref{sec:summary}, we summarize the most important points of the preceding sections and discuss some future directions for this research line.

\section{Ablation rate as a function of stellar mass and disc mass}
\label{sec:param_study}

\begin{table*}
\centering
\caption{\label{tab:stellar_params} Stellar and Wind Parameters}
\begin{tabular}{ccccccc}
\hline
$\Mstar$ ($\Msun$) & $\Rstar$ ($\Rsun$) & $\Lstar$ ($\Lsun$) & $\bar{Q}$ & $Q_\o$ & $\alpha$ & $\Mdot$ ($\msbyr$) \\
\hline
25 & 8.5 & $7.34\e4$ & 2400 & 3200 & 0.65 & $5.9\e{-8}$ \\
50 & 13.6 & $3.52\e5$ & 2100 & 2100 & 0.66 & $7.0\e{-7}$ \\
75 & 17.8 & $7.60\e5$ & 2000 & 1800 & 0.67 & $2.0\e{-6}$ \\
100 & 21.5 & $1.25\e6$ & 2000 & 1900 & 0.67 & $3.8\e{-6}$ \\
\hline
\end{tabular}
\end{table*}

In order to investigate the role of ablation in moderating accretion onto forming high mass stars, we set up simulations in much the same way as has been done in paper III.
As a brief summary of this process, we use the mass-luminosity relation derived from the Geneva stellar evolution models \citep{EksGeo12,YusHir13} with stellar radii given as a function of stellar mass by $\Rstar/\Rsun \equiv (\Mstar/\Msun)^{2/3}$.
Using this mass-luminosity relation, and particularly this mass-radius relation, assumes that the stars we are studying are main-sequence-like.
This assumption is broadly bourne out by comparison with the simulations of stellar structure in the presence of accretion carried out by \cite{HosOmu09} and \cite{HosYor10}, although these simulations show that high accretion rates of $10^{-4}\sim10^{-3}\msbyr$ can keep a massive star from contracting to its main sequence radius until it reaches $20\sim30\Msun$.
In order to generally avoid this regime in which the bloated star produces negligible or no UV feedback due to its cooler effective temperature, we only consider stars with $M_\ast \geq 25 \Msun$.

For each of the stars considered, we self-consistently calculate stellar wind launching and the radiative acceleration of circumstellar material by UV irradiation from these stars using the standard theory of line-driven acceleration first laid out by \cite{CasAbb75}, and modified for 3D velocity and radiation fields by \cite{CraOwo95}, which we have summarized in paper I.
This theory makes use of four key parameters\footnote{We use the parameterization of the \cite{CasAbb75} theory proposed by \cite{Gay95}.}, namely $\alpha$, $\delta$, $\bar{Q}$, and $Q_\o$.
Here $\alpha$ is the temperature dependent power-law index describing the distribution of spectral lines participating in the acceleration.
Meanwhile $\delta$, introduced by \cite{Abb82}, parameterizes the impact of the balance of ionization and recombination as a function of distance from the star.
The final two parameters, $\bar{Q}$ and $Q_\o$, are respectively the flux- and population-weighted enhancement of line opacity over electron scattering opacity and the maximal enhancement above which the assumed distribution of lines is exponentially truncated.
Based on the typical values cited by \cite{KudPau89}, we select a fixed value of $\delta=0.1$ for all simulations.
For the values of the remaining three parameters, we continue to consult \cite{PulSpr00}.
We summarize the stellar and wind parameters of the four stars used for simulations in this paper in table \ref{tab:stellar_params}.

In addition to the stellar parameters, we also vary the parameters of the circumstellar disc.
The disc density structure is again set by hydrostatic equilibrium in the vertical direction, $z$, with a power-law fall off in the cylindrical radial direction, $R$, of exponent $\beta$.
Taken together, this gives for the density

\begin{equation}\label{eqn:rho_d}
\rho_\d(R,z) = \rho_{\d,\o}\left(\frac{R}{R_\ast}\right)^{-\beta} e^{-z^2/(2H(R)^2)}\;,
\end{equation}
where the scale height of the Gaussian stratification is given in terms of the isothermal sound speed $c_s$, and Newton's gravitational constant $G$, to be

\begin{equation}\label{eqn:scale_height}
H(R) \equiv \frac{c_s R^{3/2}}{\sqrt{G \Mstar}}\;.
\end{equation}

In contrast to prior papers in this series, we now set the characteristic scale of disc density, $\rho_{\d,\o}$, by imposing a disc mass $M_\d$ and outer truncation radius $R_\d$, such that

\begin{equation}\label{eqn:rho0}
\rho_{\d,\o} = M_\d \frac{1}{\sqrt{2\pi^3}} \frac{1}{H(R_\ast)} \frac{1}{R_\d^2-R_\ast^2}\;.
\end{equation}
By comparison with observations \citep[see][for a review]{BeldeW16} and numerical simulations \citep[e.g.][]{KlaPud16,RosKru16,KuiHos18} we use 1000 au as a fixed value of $R_\d$ to normalize $\rho_\mathrm{d,o}$, although the computational domain of the simulations we carry out here extends to only 20 $\Rstar$.
By similar comparisons, we select 3 values of $M_\d$, specifically 10, 20, and 30 $M_\odot$.

From these initial conditions, we run 2D, azimuthally symmetric, isothermal simulations on a spherical $\{r,\theta\}$ grid using the MHD code Pluto \citep{MigBod07,MigZan12}.
All simulations use evenly spaced grids of 257 cells in the polar direction from pole-to-pole and stretched grids of 256 cells in the radial direction, with each cell 3\% larger than its nearest neighbour in the star-ward direction.
We impose impermeable boundaries at both poles, and only allow material to pass through the stellar photosphere boundary at speeds less than $c_s$.
As we do not want to allow feedback from the outer boundary onto the simulation volume, we only allow material to exit the simulation through this boundary.

It is important to note that the geometry of accretion in the presence of ablation is not known, and may be altered by the presence of shear from the ablation layers even at substantial distances from the stellar surface.
Indeed, since these simulations are among the first to probe the inner accretion discs of massive (proto)stars, it is not even clear how accretion should be expected to proceed onto massive (proto)stars if ablation was omitted.
Turning to observations, \cite{AbaOud17} have recently probed this region with spectropolarimetry, with the conclusion that accretion onto stars earlier than about B7-B8 does not produce the same spectral features as the more well studied T Tauri stars.
This is further inferred to mean that these earlier type stars do not accrete with the same geometry as their lower mass analogues.
However, these observations do not by themselves tell us what the geometry of this accretion is, underscoring the necessity of additional studies to understand this region.

Therefore, given this current general lack of additional constraints, we continue to omit the replenishment of disc material by accretion, and as such the accretion onto the (proto)star.
This omission means that all simulations will completely eject their initial disc material given enough time, regardless of whether the rate of ablation exceeds the accretion rates expected for very massive star formation \citep[$10^{-4} \sim 10^{-3} \msbyr$, e.g.][]{WolCas87}.
As papers I and III showed, this ejection of disc material will proceed from the stellar surface outwards.
Increasing the initial reservoir of mass in the disc, however, delays the opening of this gap between the stellar photosphere and disc edge.
Therefore, in addition to exploring what the effects of increased density in the regions being ablated might be, we can also view the simulations with increased disc mass to be a proxy for the replenishment of disc material by accretion. 

\begin{figure*}
\centering
\begin{subfigure}{0.48\textwidth}
\includegraphics[width=\textwidth]{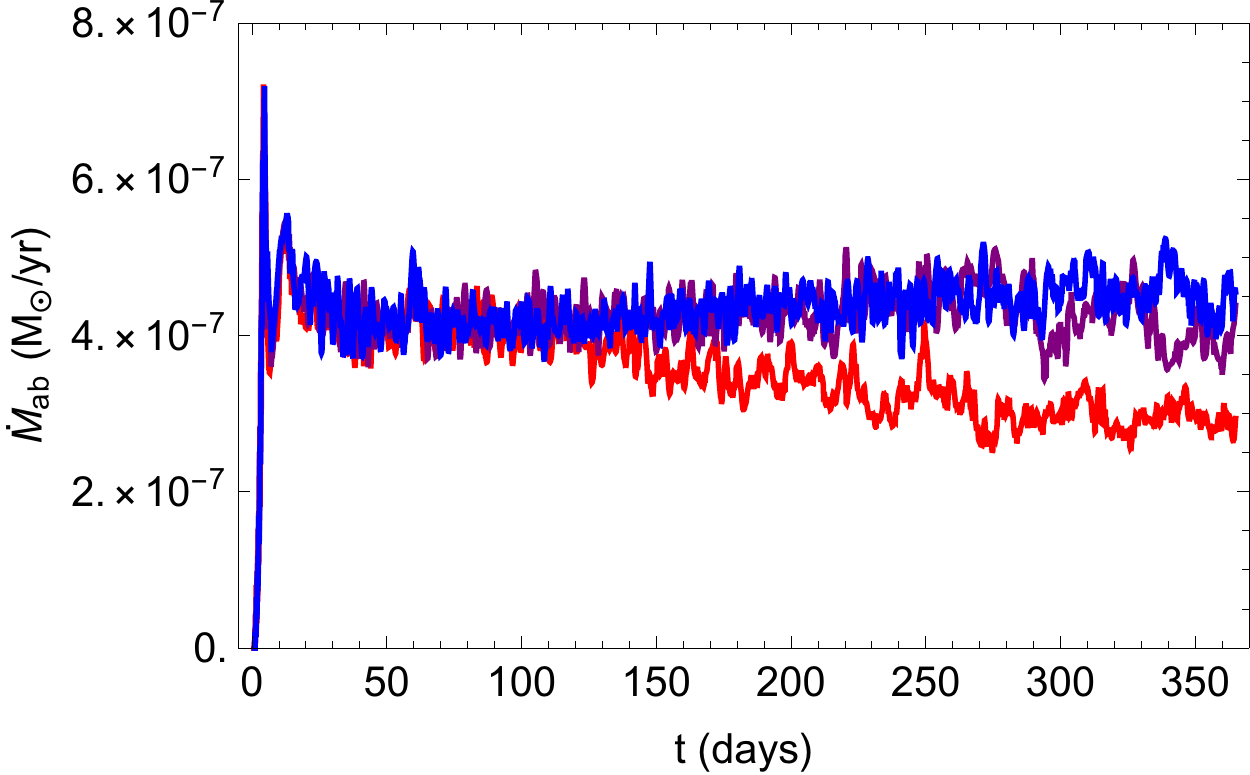}
\caption{$\Mstar = 25 \Msun$}
\end{subfigure}
\hspace{5 pt}
\begin{subfigure}{0.48\textwidth}
\includegraphics[width=\textwidth]{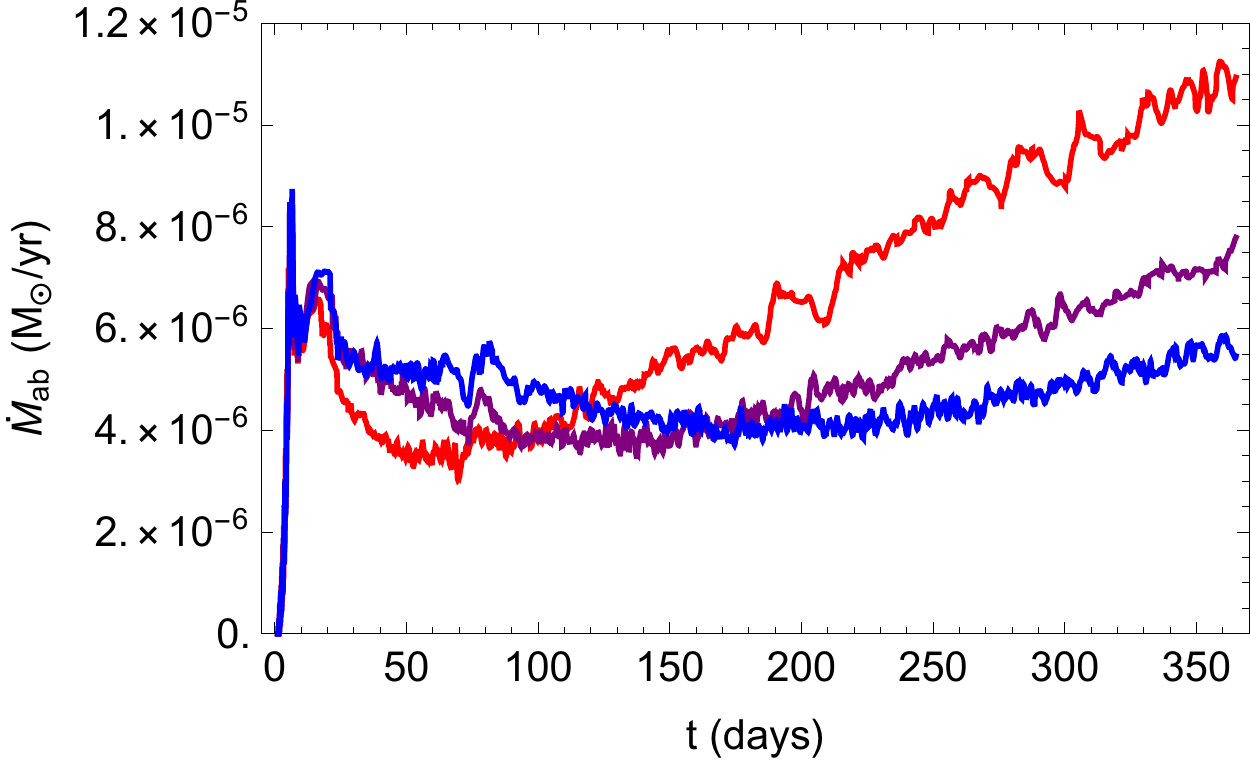}
\caption{$\Mstar = 50 \Msun$}
\end{subfigure}
\vspace{10 pt}

\begin{subfigure}{0.48\textwidth}
\includegraphics[width=\textwidth]{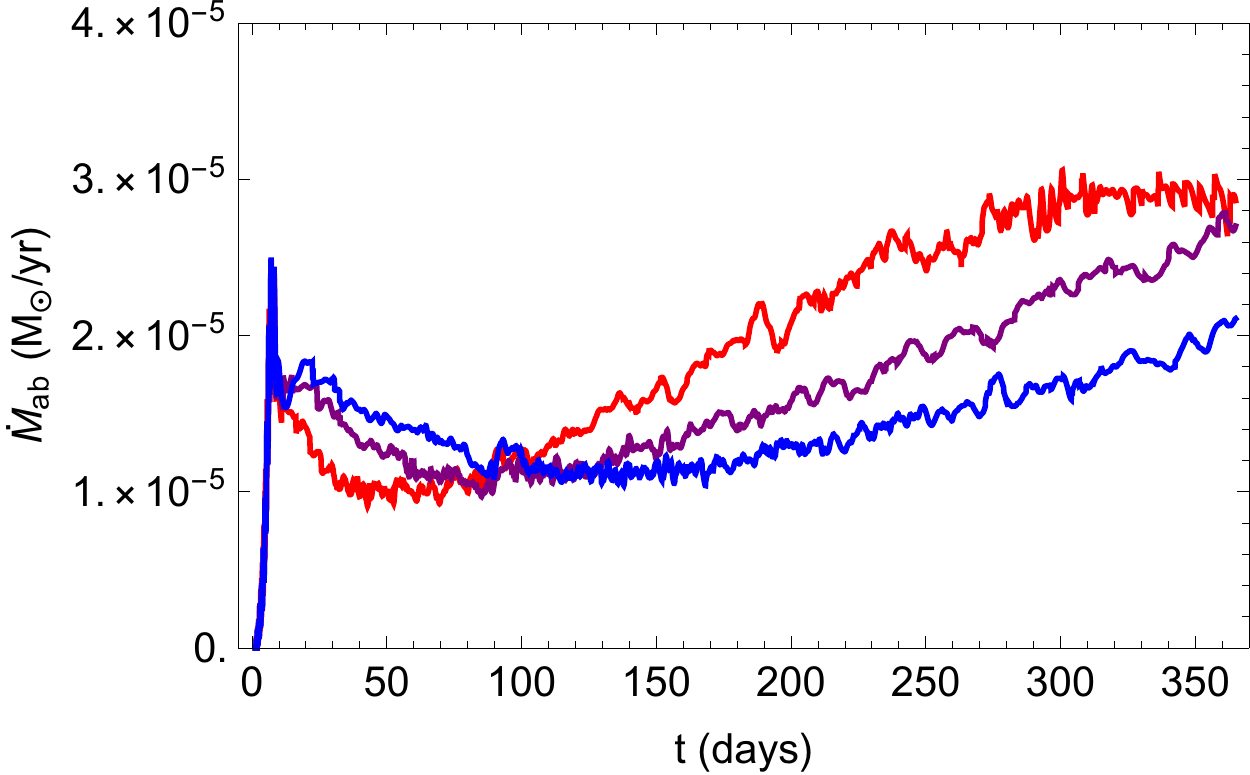}
\caption{$\Mstar = 75 \Msun$}
\end{subfigure}
\hspace{5 pt}
\begin{subfigure}{0.48\textwidth}
\includegraphics[width=\textwidth]{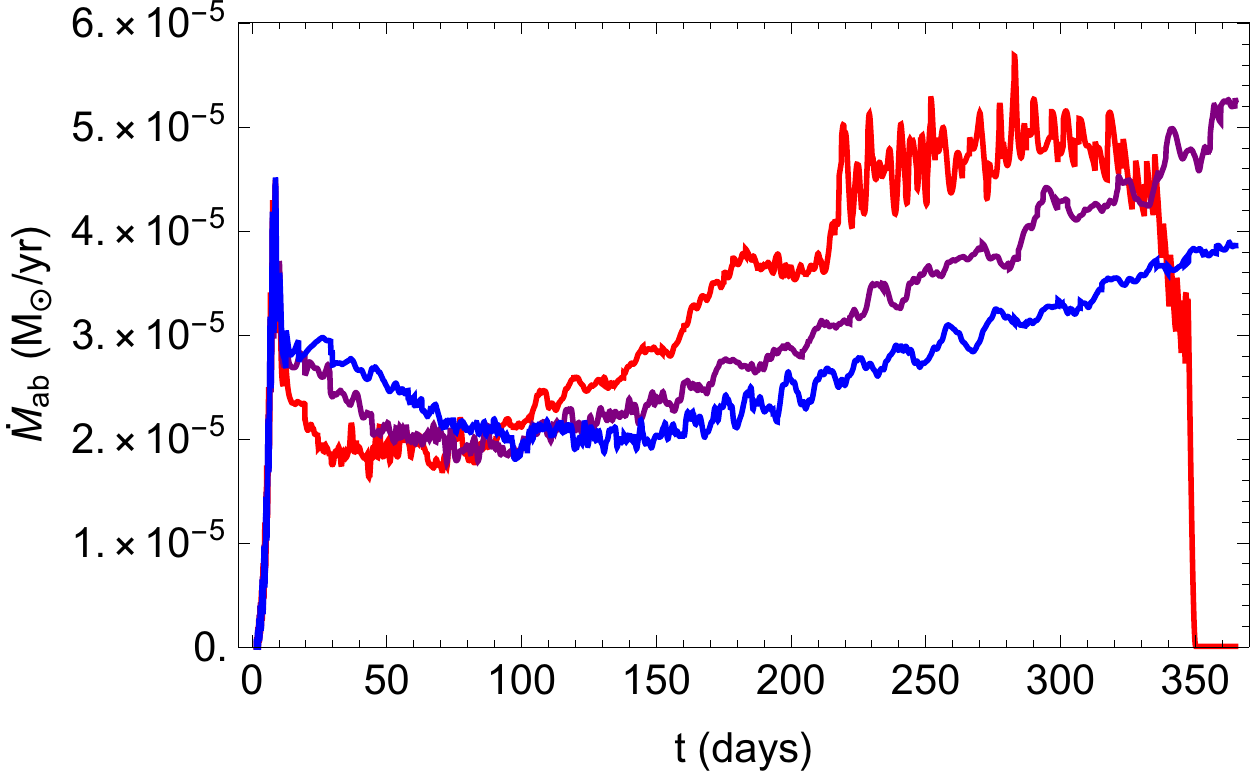}
\caption{$\Mstar = 100 \Msun$}
\end{subfigure}
\caption{\label{fig:ablation_comparison} Disc ablation rate normalized by spherically symmetric wind mass loss rate for 25 (top, left), 50 (top, right), 75 (bottom, left), and 100 (bottom, right) solar mass stars. In each panel the red, purple, and blue curves respectively denote the results from simulations with initial disc densities representing 10, 20, and 30 solar mass discs respectively (see the density normalization definition in equation \ref{eqn:rho0}).}
\end{figure*}

When we analyse the results of these simulations, the main quantity of interest is ablation rate, $\dot{M}_\mathrm{ab}$, calculated by taking the mass flux at supersonic velocities through the outer radial boundary minus the spherically-symmetric wind mass loss rate, $\dot{M}_\mathrm{wind}$.
Recall from paper III that the choice to only use the rate at which mass supersonically leaves the simulation is based on the presence of sub-sonic waves set up in the disc by the relaxation from the initial conditions.
As shown by paper III, these waves average out over time, but can nevertheless contaminate the measurement of ablation rate as they occur in very dense material deep in the disc.
Figure \ref{fig:ablation_comparison} plots the computed ablation rate as a function of time in solar masses per year for all the simulations.

Immediately evident in examining this figure is the approximately two order of magnitude spread in the rate of ablation between the different stars considered.
This implies that even relatively small changes in stellar mass lead to fairly severe changes in the impact of line-driven ablation on the circumstellar accretion disc, due to the steeply increasing luminosity of the central star.
Indeed, this effect is so strong that for the lowest disc mass to mass loss rate ratio, namely the $10 \Msun$ disc around a $100 \Msun$ star, ablation fully remove the disc from the simulation volume before the simulation completes, hence the precipitous drop off in ablation rate at around 350 days in this simulation (represented by the red curve in the bottom right hand panel of figure \ref{fig:ablation_comparison}).

From figure \ref{fig:ablation_comparison} it is also worth noting that the simulations with the highest ratio of disc mass to wind mass loss rate, i.e. the simulations of 20 and $30 \Msun$ discs around the $25 \Msun$ star, have the least time variability in the ablation rate.
This same trend of increasing disc mass leading to a more steady rate of disc ablation can be seen as well for the higher mass stars, although the increased strength of line-driven acceleration from these  stars mean that a $30 \Msun$ disc does not provide a sufficient mass reservoir to allow such a steady state to develop before the disc is separated from the star.

In continuing to discuss these results, we draw the readers attention to the general trend that, with the exception of the simulations with a $25 \Msun$ star, the ablation rate initially decreases, then later increases in time.
We interpret this as the competition of two effects.
The first of these is a decreasing efficiency of ablation with increasing distance from the star.
As shown particularly vividly by papers I and II, photons arriving to the disc from near the stellar limb, or from off-star directions as was the case in paper II, interact more effectively with the Keplerian shear of the disc than photons arriving to the disc nearly radially from the star.
As time goes on, an increasingly large gap is opened between the disc and star and the star becomes more point-like and thus less effective at driving ablation, hence the initial decrease in ablation rate.
However, competing with this is the increased strength of line-acceleration on lower density material.
Since ablation tends to strip the surface layers off the disc leaving the remainder of the disc largely un-perturbed, ablation is exposing  decreasing density disc material at the inner disc rim closest to the star that was previously shielded by the inner disc.
The inverse dependence of line-acceleration on density means that this lower density material is easier to accelerate.
Additionally, the wind has reached its terminal velocity of several thousand km s$^{-1}$ at these distances, and the shearing of the wind against the disc may also contribute to entraining some disc material.
Thus, the decreasing density away from the star leads to the eventual late time increase in ablation rate.

Returning to the simulations with a $25\Msun$ star shows that, while they initially appear to behave differently, they are also consistent with this picture.
The simulation with a $10\Msun$ disc around a $25 \Msun$ star opens only a small gap of order $\Rstar$ between the stellar photosphere and the inner disc rim over the whole year of simulation time. 
The corresponding simulation with a $50\Msun$ star and the same $10\Msun$ disc mass opens a comparable gap in only $\sim$50 days, at which point its mass ablation rate has not yet started increasing, as is the case for the $25 \Msun$ star with a $10\Msun$ disc.
The other two simulations with a $25 \Msun$ star have opened even smaller gaps, such that the downturn in ablation rate is not particularly substantial over the one year simulation duration.

\begin{figure}
\centering
\includegraphics[width=0.49\textwidth]{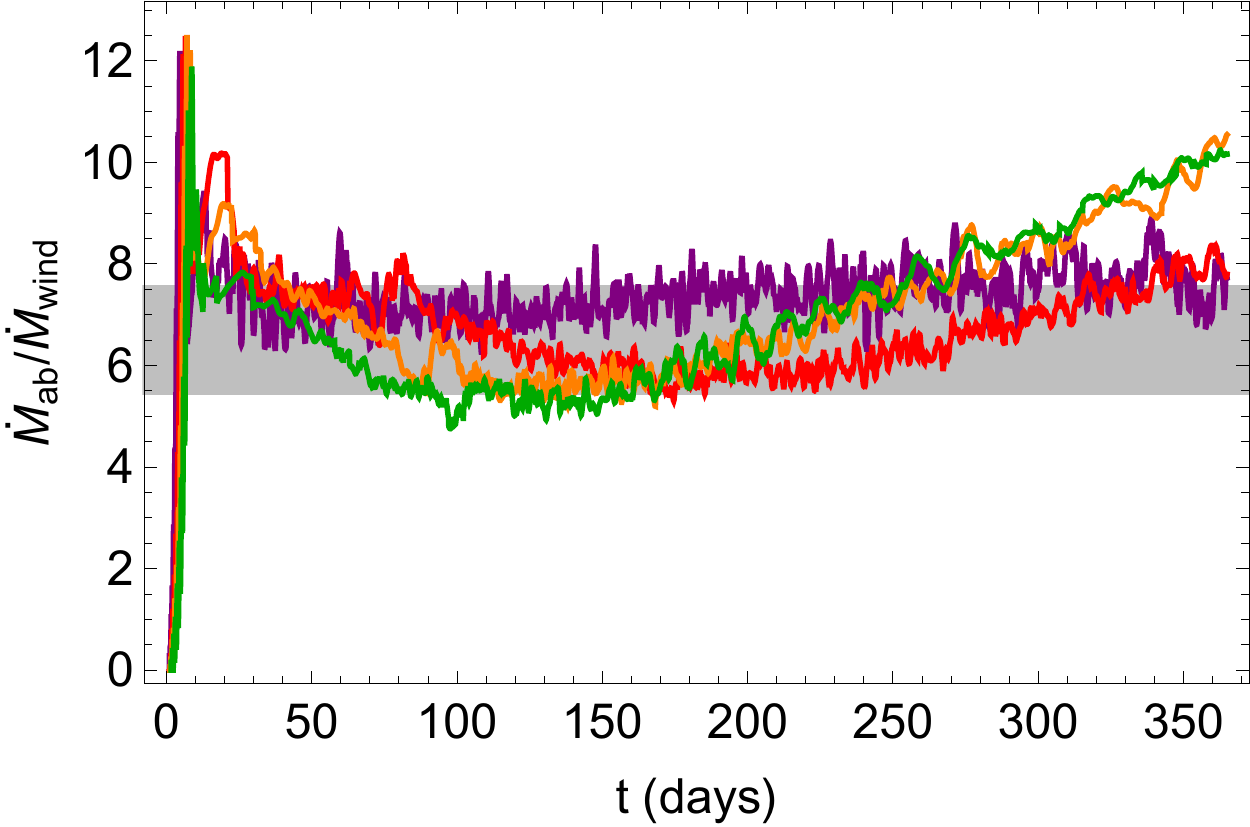}
\caption{\label{fig:Mdisk30_rate_overplot}An overplot of the normalized ablation rates from all the simulations beginning from 30 solar mass discs, with the results of the simulations with 25, 50, 75, and 100 solar mass stars denoted by the purple, red, orange, and green curves respectively. The shaded bar at $\dot{M}_\mathrm{ab}/\dot{M}_\mathrm{wind}=6.5\pm 1$ is put in by eye to call out the overall uniformity of behavior between different simulations.}
\end{figure}

Given both the lower degree of time variability, and the increased resemblance to what we would expect in the case of replenishment of the disc by accretion, general quantitative conclusions about ablation rate should be made from the simulations with the highest disc mass to mass loss rate ratios.
To this end, we replot the ablation rates from all the simulations including $30 \Msun$ discs in figure \ref{fig:Mdisk30_rate_overplot}, now in units of $\dot{M}_\mathrm{wind}$ for each of the stars.
When we plot the simulation results in this way, it becomes evident that the general behavior of line-driven ablation for discs of star forming densities can again be parameterized as a fixed factor enhancement over the stellar wind mass loss rate, as was done in paper I.
In order to guide the eye, figure \ref{fig:Mdisk30_rate_overplot} includes a shaded bar at $\dot{M}_\mathrm{ab}/\dot{M}_\mathrm{wind}=6.5\pm 1$.
As the ablation rate for all simulations falls more or less in this range over the duration of the simulations, we argue that ablation rate can be generally parameterized by a fixed factor enhancement over the spherically symmetric stellar mass loss rate.
%\revision{
Further examination of figure \ref{fig:ab_rate} shows that this enhancement factor seems to be relatively independent of the density of the disc as well, with all the resulting ablation rates falling in approximately the same range for each star considered.
However, it is difficult to make conclusive statements on this dependence as decreasing the disc density results in the ablation rate being less constant in time and therefore more difficult to quantify as the disc separates from the star.
%}
Thus we proceed with the analyses in the remainder of the paper assuming $\dot{M}_\mathrm{ab}=6.5\;\dot{M}_\mathrm{wind}$.

%\revision{
In order to begin disentangling why it is possible for $\dot{M}_\mathrm{ab}$ to exceed $\dot{M}_\mathrm{wind}$ it is of key importance to recall that the ablation of disc material is not driven by radially streaming photons, but rather by photons arriving to the disc from non-radial directions causing the roughly radial alignment of the ablation layer to not be unfavorable to the initiation of disc mass loss.
The importance of this non-radial photon flux is highlighted by the fact, as shown in papers I and III, that most of the disc mass loss originates near the inner edge of the disc, e.g. where the star subtends the largest solid angle on the sky.
Additionally, since the disc is initially in gravito-centrifugal equilibrium, a smaller force is needed to launch the ablative mass loss than to launch the wind, allowing ablation to be launched at higher densities than the wind.
Finally, the flaring of the isothermal disc we simulate here also allows the wind to run into the disc, thereby entraining some additional material into the ablation layer as it flows away from the star.
%}

%\revision{
At this point, we can also compare to the models run by \cite{DrePro98} and \cite{OudPro98} of disc ablation around a 10 $\Msun$ star.
The simulation set-ups in this paper and in \cite{DrePro98} and \cite{OudPro98} are quite similar, with the exception that we do not account for contributions to the line-driven acceleration from stellar radiation reprocessed by the disc, which these prior works do.
Additionally, \cite{DrePro98} and \cite{OudPro98} fix the density along the disc midplane in order to attempt to account for disc replenishment by accretion.
Comparing the results of the simulations shows that \cite{DrePro98} and \cite{OudPro98} find a much thicker ablation layer than we do, likely the result of the reprocessed stellar radiation in the disc acting perpendicular to the ablation layer.
Additionally, \cite{DrePro98} finds an enhancement factor of 3, rather than the 6.5 that we find\footnote{\cite{OudPro98} does not present this quantity.}.
Coupled with the lower enhancement factors we found in paper I, this suggests a weak dependence of ablation enhancement on either stellar or disc parameters, which we posit may be due to the weaker line-driving and decreased luminosity for later spectral types.
Since the spread in enhancement factors is small, to first order the choice of a fixed enhancement factor is sufficiently good for us to proceed with the analyses in this paper. 
Nevertheless, future work should investigate this difference further, as well as including stellar radiation reprocessed by the circumstellar disc in order to confirm that this omission is indeed the origin of the thinner ablation layer we find.
%}

Before undertaking the analyses in section \ref{sec:upper_mass_lim}, it is worth pointing out that $\dot{M}_\mathrm{ab}=6.5\;\dot{M}_\mathrm{wind}$ is probably a conservative interpretation of the results of these simulations.
As we have already argued, the simulation with a 25$\Msun$ star and a 30$\Msun$ disc is the one most like what we would expect from simulations with self-consistent disc replenishment by accretion.
While the quite time independent ablation rate of this simulation reinforces the interpretation that ablation proceeds at a fixed enhancement over wind mass loss rate, it also seems more consistent with this enhancement being $\sim 8$.
However, future simulations which treat accretion and ablation side-by-side are required to constrain the exact value of this enhancement factor.
%\revision{
Such simulations will also allow for a more detailed investigation of the dependence of the ablation rate on disc density, as the disc density can be more freely varied without the disc blowing away, especially for the more luminous stars considered.
%}
In the absence of such simulations, we argue that the conservative choice is the most appropriate one, and use this enhancement factor to make some general conclusions about the importance of ablation in the formation of very massive stars.

\section{From ablation rate to efficiency of accretion}
\label{sec:upper_mass_lim}

Given that we can predict the strength of line-driven disc ablation based on the analytic wind mass loss rate, itself a function of stellar parameters, a natural next step is to predict how important ablation should be on the scale of accretion for the formation of stars of various masses.
We first examine this scaling at solar metallicity in the following subsection, \ref{sec:solarZ}, before extending the discussion in subsection \ref{sec:Z_scaling} to address the expected scaling to lower metallicity environments.

\subsection{Solar Metallicity}
\label{sec:solarZ}

\begin{figure}
\centering
\includegraphics[width=0.49\textwidth]{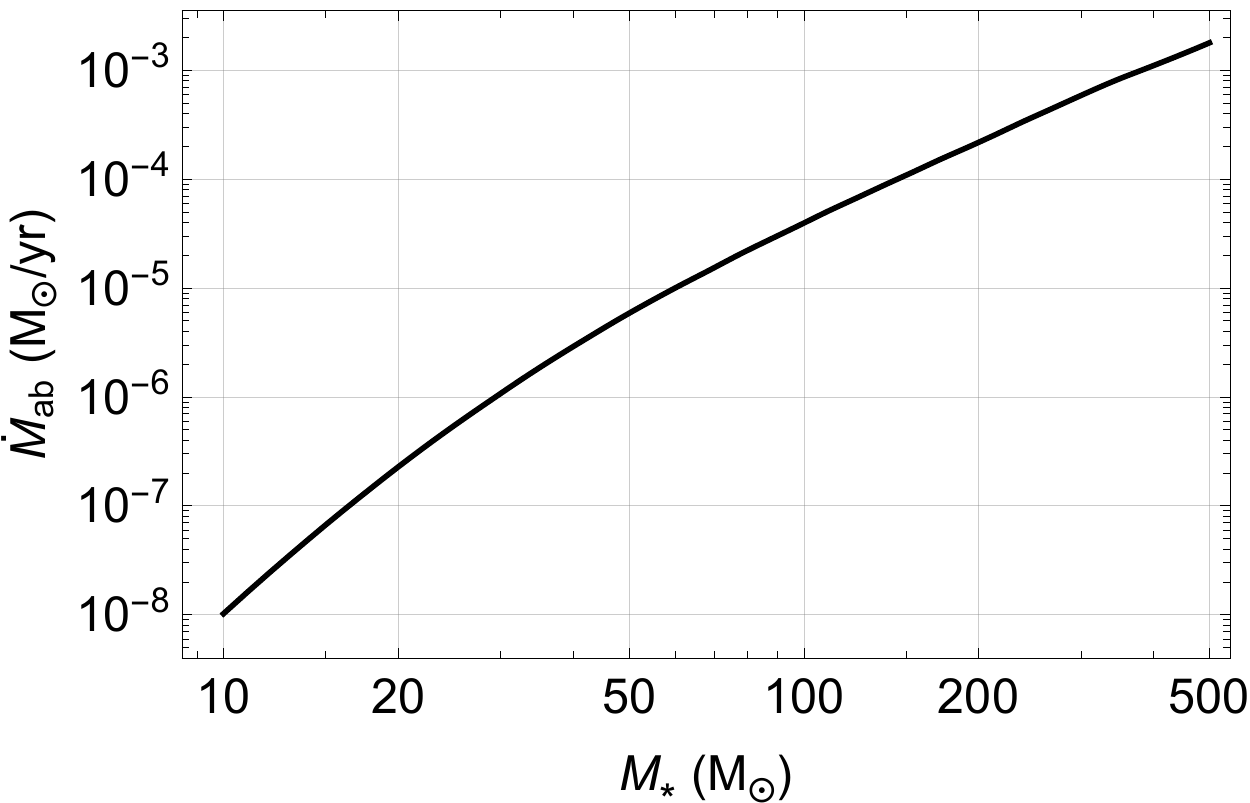}
\caption{\label{fig:ab_rate} Predicted ablation rate in solar masses per year as a function of stellar mass.}
\end{figure}

For these analyses, we use the full, analytic mass loss prescriptions described by \cite{KudPau89} for $\dot{M}_\mathrm{wind}$, and take $\dot{M}_\mathrm{ab}=6.5\;\dot{M}_\mathrm{wind}$ as argued in section \ref{sec:param_study}.
Into the expression for $\dot{M}_\mathrm{wind}$, we plug the ZAMS mass-luminosity relation from the Geneva stellar evolution tracks \citep{EksGeo12,YusHir13} and the mass-radius relation $\Rstar/\Rsun = (\Mstar/\Msun)^{2/3}$.
For stars more massive than those included in the ZAMS Geneva tabulations, we extrapolate using $\Lstar \propto \Mstar^3(1-\Gamma)^4$, where $\Gamma \equiv \kappa_\mathrm{e} \Lstar/4 \pi G \Mstar c$ is the electron scattering Eddington parameter, here used to prevent us from considering stars where radiation pressure on the basal electron scattering gas opacity exceeds the gravitational binding of the star.
For simplicity, we fix the wind parameters to be $\bar{Q}=Q_\o=2000$, $\alpha=0.67$, and $\delta=0.1$.
Note that, as shown by \cite{PulSpr00}, the assumption that $\bar{Q}=Q_\o$ is not valid for stellar effective temperature below about 30 kK.
However, the stellar effective temperature predicted from the scaling relations we use here exceeds this threshold for stars of a few tens of solar masses, making $\bar{Q}=Q_\o$ a good assumption in the parameter range considered (see also table \ref{tab:stellar_params}).
Using these scalings and assumed parameters, figure \ref{fig:ab_rate} plots the derived $\dot{M}_\mathrm{ab}$ as a function of stellar mass in solar masses per year.

Immediately worth noting from figure \ref{fig:ab_rate} is the five orders of magnitude that the rate of ablation spans over the range of stellar masses considered.
This underscores the point brought up in section \ref{sec:param_study} that ablation can have vastly different impacts on disc structure and dynamics for comparably small changes in the stellar mass.
For instance, even the densest discs considered in section \ref{sec:param_study} are significantly disrupted by the ablation feedback of a $\Mstar \geq 50 \Msun$ star, while the ablation driven by a $25 \Msun$ star barely impacts the disc at all.

In order to assess whether the absolute magnitude of ablation shown in figure \ref{fig:ab_rate} in general makes it an important feedback mechanism or not, one potential method is to compare it to the accretion rates predicted by theories of massive star formation.
To this end, we select two such theories, namely Competitive Accretion \citep{BonBat01} and Turbulent Core Accretion \citep{McKTan03}.
Comparisons of these two theories in general suggest that Competitive Accretion predicts lower accretion rates \citep{TanBel14}, so we choose to first compare the strength of ablation against it.
However, the parameterization of accretion rate presented by \cite{BonBat01} as a function of variables which are not in general known, such as the relative speed of stars and gas and the position in the cluster where the star is forming, makes such a comparison complicated.
Therefore, we base this comparison on the average rate of accretion proposed for Competative Accretion by \cite{TanBel14}, specifically their equation 16,

\begin{multline}
\langle \dot{M}_\mathrm{acc,comp} \rangle = 1.46\e{-4} \msbyr \\
\times \frac{\epsilon_\mathrm{ff}}{0.1}\frac{M_{\ast,\mathrm{f}}}{50 \Msun}\frac{0.5}{\epsilon_\mathrm{cl}}\left(\frac{1000 \Msun}{M_\mathrm{cl}}\right)^{1/4}\Sigma_\mathrm{cl}^{3/4}\;,
\end{multline}
where $\epsilon_\mathrm{ff}$ is the star formation efficiency per free-fall time, $\epsilon_\mathrm{cl}$ is the total star formation efficiency from the star forming clump, $\Sigma_\mathrm{cl}$ is the characteristic surface density of the star forming clump, and $M_{\ast,\mathrm{f}}$ is the final mass of the star.
Taking as fiducial parameters $\epsilon_\mathrm{ff}=0.1$, $\epsilon_\mathrm{cl}=0.5$, and $\Sigma_\mathrm{cl}=1$ g cm$^{-2}$, this equation predicts that the average accretion to form a 150$\Msun$ star from a $2\e4\Msun$ clump is $2\e{-4}\msbyr$, comparable to the ablation rate we predict at $\Mstar=150\Msun$, making line-driven ablation a key feedback effect to consider in such a system as it alone may be strong enough to shut off accretion.

Comparing now to Turbulent Core Accretion, \cite{McKTan03} argue for a rate of accretion given by their equation 41 to be

\begin{equation}
\dot{M}_\mathrm{acc,core} = 4.6\e{-4} \left(\frac{M_{\ast,\mathrm{f}}}{30 \Msun}\right)^{3/4} \Sigma_\mathrm{cl}^{3/4}\left(\frac{M_\ast}{M_{\ast,\mathrm{f}}}\right)^{1/2}\msbyr\;.
\end{equation}
While this is a function of stellar mass, the scaling here is much shallower than the scaling of ablation rate with stellar mass, such that ablation will be the most important when $M_\ast = M_{\ast,\mathrm{f}}$.
Thus, for the same $150\Msun$ star again forming in an environment with $\Sigma_\mathrm{cl}=1$ g cm$^{-2}$, Turbulent Core Accretion predicts an accretion rate of $1.5\e{-3}\msbyr$.
While this is a good deal higher than the rate predicted by Competitive Accretion, ablation still can remove $5\sim10\%$ of such an accretion flow, suggesting that it could still play an important role in moderating accretion onto a forming star under Turbulent Core Accretion.

\begin{figure}
\centering
\includegraphics[width=0.49\textwidth]{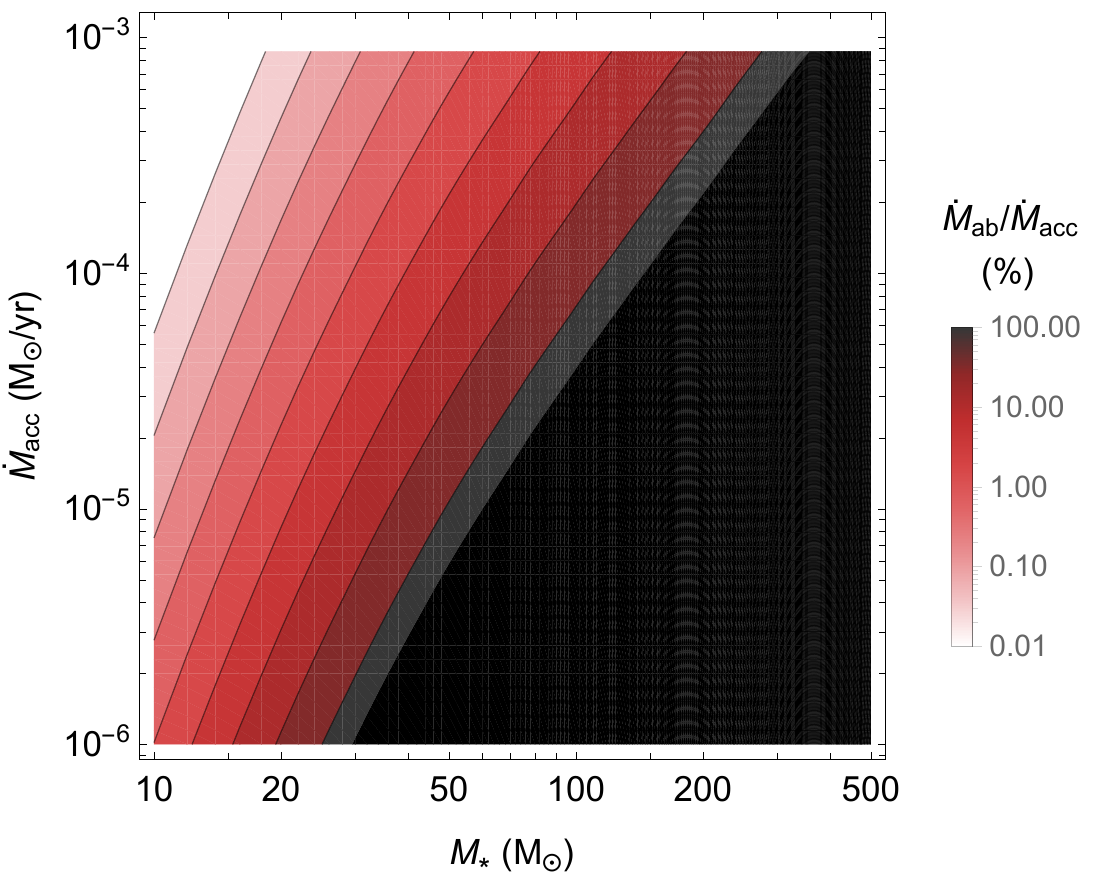}
\caption{\label{fig:mass_lim} The percentage of accretion predicted to be removed by ablation as a function of stellar mass and accretion rate. The black region in the lower right denotes the parameter space where ablation rate is predicted to exceed accretion rate, such that stars can not form here by steady disc accretion.}
\end{figure}

We can also assess the strength of ablation more generally by  plotting the percentage of accretion that would be predicted to be removed by ablation for mass accretion rates from $10^{-6}-10^{-3}\; \msbyr$, as is done in figure \ref{fig:mass_lim}.
Here again the variation in impact of ablation with stellar mass is quite apparent.
Now, instead of needing to infer it however, we can directly see the excluded region in the lower right of the plot (black), where we predict stars can not form by steady disc accretion, as ablation rate alone exceeds accretion rate, and thus would be expected to detach the star from its disc.
Note that highly optically thick accretion streams, blobs/fragments of gas, and of course stellar mergers can circumvent this barrier, as they locally result in accretion conditions in the upper left of figure \ref{fig:mass_lim} even when the global, time averaged accretion properties may fall in this excluded region to the bottom right.
In the absence of such impulsive accretion events, however, this region shows that the formation of stars of $50 \Msun$ is excluded for accretion rates below about $3\e{-6}\msbyr$ and even the 50\% increase in mass to a $75 \Msun$ star requires a minimum accretion rate a factor of three higher, i.e. $\sim 10^{-5} \msbyr$.
To form the most massive stars stars in the Milky Way \citep[e.g. the central stars in NGC 3603 whose individual masses are inferred to exceed $120 \Msun$,][]{MelMas08}, the constraints are yet more stringent, requiring mass accretion rates of order $\sim 10^{-4} \msbyr$ or higher.

In interpreting the remainder of figure \ref{fig:mass_lim}, it is important to remember that by isolating ablation we have neglected a variety of feedback mechanisms predominantly taking place on larger scales.
As shown by, for instance, \cite{KlaPud16}, \cite{RosKru16}, \cite{KuiTur16}, \cite{TanTan17}, and \cite{KuiHos18}, these other feedback mechanisms can already substantially reduce the mass able to make it out of a pre-stellar core and into a massive (proto)star.
In some simulations, these other feedback mechanisms are even already sufficient to shut off accretion completely, regardless of the mass reservoir available \citep[see specifically the discussion in][]{KuiHos18}.
Therefore, the percentage of accretion deflected by ablation as predicted in figure \ref{fig:mass_lim}, or the absolute magnitude of ablation rate as presented in figure \ref{fig:ab_rate}, should be interpreted as further reducing the accretion flow beyond what is already predicted at larger radii.
Indeed, given that line-driven ablation operates on such a small scale compared to these other feedback mechanisms, this percentage of the accretion flow could be reinjected as a boundary condition in larger scale simulations, thus naturally allowing ablation to be analytically included in future work at these larger scales.

Moving to the low mass end of figures \ref{fig:ab_rate} and \ref{fig:mass_lim}, it is important to recall that, although the Kelvin-Helmholtz time can be shorter than the accretion time for massive stars \citep{Kah74}, computations of stellar structure with an accretion boundary condition show that very high accretion rates can keep stars bloated during the early phases of their formation, perhaps even up to $\Mstar \sim 25 \Msun$ for $\dot{M}_\mathrm{acc}\geq10^{-3} \msbyr$ \citep{HosOmu09,HosYor10}.
Indeed, as we discussed in section \ref{sec:param_study}, it is this effect that motivated the choice to compute models only for stars of $\Mstar \geq 25 \Msun$, as protostars bloated by accretion will not emit sufficient radiation in the UV to initiate line-driven stellar winds, and thus will also not initiate disc ablation.
This shutting off of ablation would predominantly affect the upper left corner of figure \ref{fig:mass_lim} where high accretion rates and relatively lower stellar masses meet, making any conclusions drawn here likely unreliable.
However, examining this corner of the figure shows that ablation amounts to $0.1\%$ or less of accretion in this parameter range, thereby allowing us to safely ignore bloating of the star for studies of ablation, as protostellar bloating only occurs in parameter ranges where ablation is already dynamically unimportant.

Finally, before discussing the scaling of ablation with metallicity, it is perhaps worth recalling that the analyses here are based on only one scaling of mass, luminosity, and radius, and that we have fixed the stellar wind parameters for simplicity.
Changing the scaling of stellar parameters, or relaxing the approximation of fixed wind parameters could lead to order unity adjustments in the specific values of ablation rate presented here, and attendant changes in the specific value of the upper mass limit imposed by ablation.
Nevertheless, given the strong scaling of mass loss rate with stellar mass, the overall, general trends presented here are robust.

\subsection{Scaling with metallicity}
\label{sec:Z_scaling}

\begin{figure}
\centering
\includegraphics[width=0.49\textwidth]{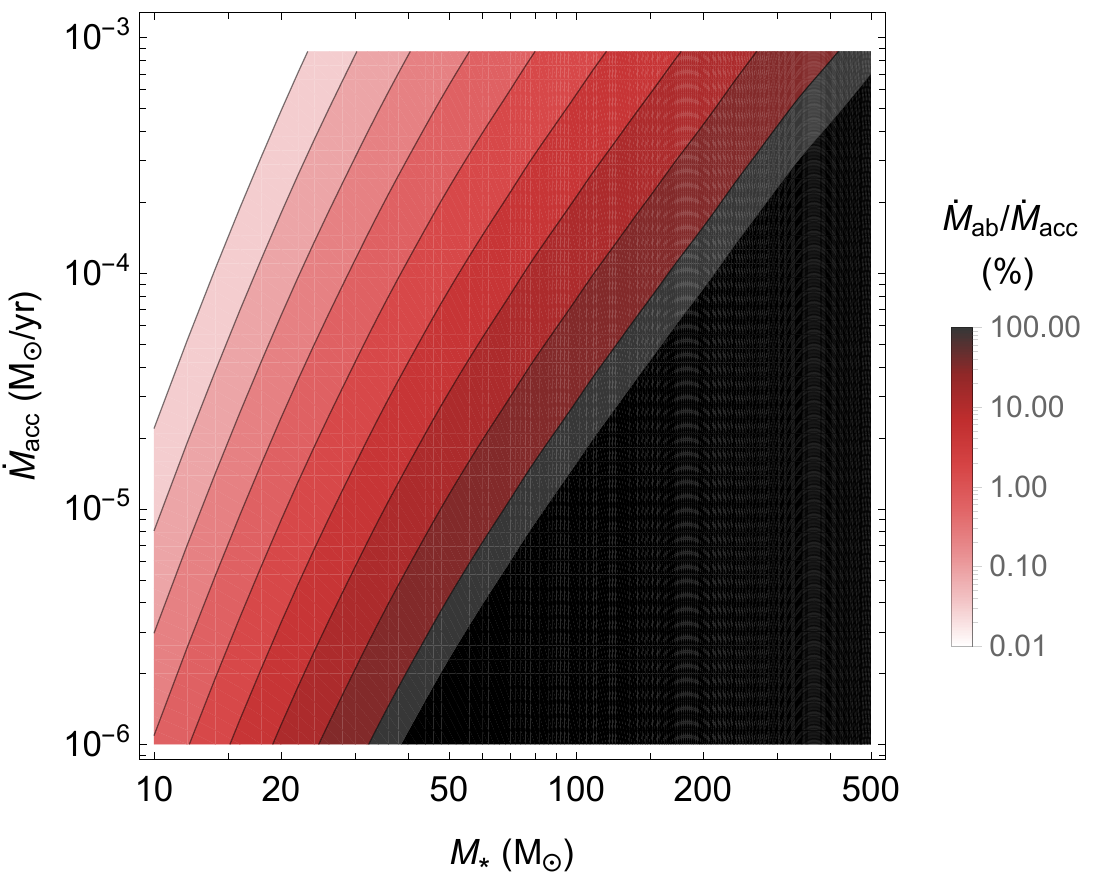}
\caption{\label{fig:mass_lim_SMC} The percentage of accretion predicted to be removed by ablation as a function of stellar mass and accretion rate at an SMC metallicity of 0.2$Z_\odot$. The black region in the lower right denotes the parameter space where ablation rate is predicted to exceed accretion rate, such that stars can not form here.}
\end{figure}

As a final aspect of the discussion in this paper, we address the scaling of ablation rate with metallicity, $Z$.
As discussed by \cite{Gay95}, the scaling of $\bar{Q}$ with $Z$ is very simple, i.e. $\bar{Q}\propto Z$.
Since the assumption $Q_\o = \bar{Q}$ is reasonable in the parameter range considered, we can then simply vary this constant to get a sense of the effects of varying metallicity.
While the overall scaling of mass loss rate with stellar parameters as laid out by \cite{KudPau89} is non-trivial, $\bar{Q}$, and thus $Z$, enters the relation only peripherally, such that the total scaling of mass loss rate with metallicity is $\dot{M}_\mathrm{wind} \propto Z^{(1-\alpha)/\alpha}$.
Continuing with the fixed $\alpha = 0.67$ used in the prior subsection, then $\dot{M}_\mathrm{wind} \propto Z^{0.5}$.
As an example, figure \ref{fig:mass_lim_SMC} shows how ablation rate would compare to accretion rate at $Z = 1/5 Z_\odot$, comparable to the conditions of the Small Magellanic Cloud (SMC).
For this factor five reduction in metallicity, mass loss driven by line-acceleration is reduced to approximately 45\% of the rate at solar metallicity, leading to an attendant increase of the upper mass limit imposed by ablation by a factor $1.5\sim 2$.

\begin{figure}
\centering
\includegraphics[width=0.49\textwidth]{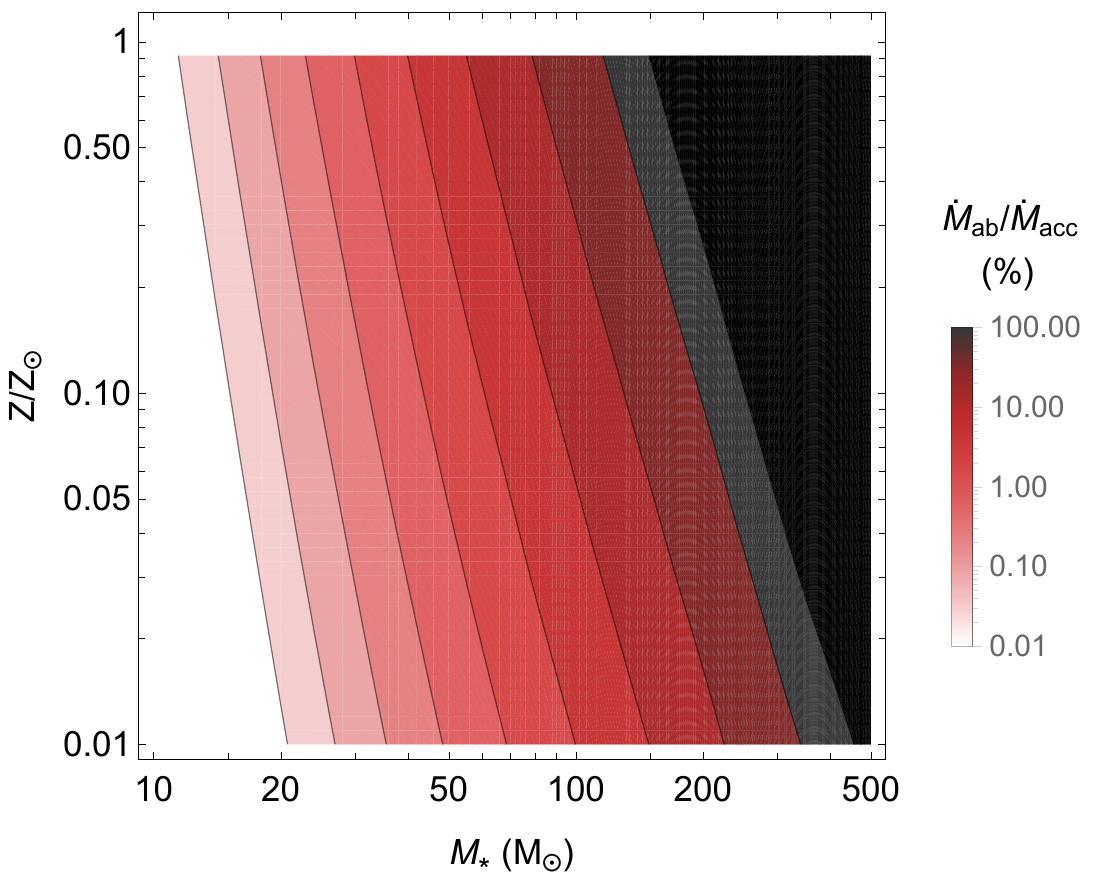}
\caption{\label{fig:mass_lim_Zdep} The percentage of a $10^{-4}\msbyr$ accretion predicted to be removed by ablation as a function of stellar mass and metallicity. The black region in the upper right denotes the parameter space where ablation rate is predicted to exceed accretion rate, such that stars can not form here.}
\end{figure}

Under the assumption that the intrinsic rate of accretion in the absence of feedback is not substantially altered by the decrease of metallicity, we select an accretion rate of $10^{-4}\msbyr$ and plot the percent of this that would be expected to be repelled by ablation as a function of $Z$ and $\Mstar$ in figure \ref{fig:mass_lim_Zdep}.
Here we directly see the monotonically decreasing strength of disc mass loss from line-driven ablation with decreased metallicity, and the attendant rise in the upper mass limit imposed by this feedback.

Note that the results presented in this subsection are based on the assumption that the enhancement of ablation over wind mass loss is not itself a function of metallicity.
Constraining the dependence of the enhancement of $\dot{M}_\mathrm{ab}$ over $\dot{M}_\mathrm{wind}$ on $Z$ would require further simulations, likely accounting for non-isothermal effects as well as self-consistently treating accretion and ablation in the same simulation. 
This is because reduced metallicity should cause a reduced efficiency of radiative cooling, thus altering both the disc structure and the geometry of accretion onto the central protostar.

Finally, as in the prior sections, it is worth bearing in mind that this analysis ignores all other feedback processes.
Additionally, not all feedback mechanisms show this decrease in efficacy with decreasing metallicity.
Take, for instance, photoevaporation.
Comparing the results of \cite{HosHir16} to those of \cite{KuiHos18} shows that photoevaporation becomes the main radiative feedback from massive stars at low metallicity.
This is supported by the models presented in \cite{TanTan18} which argue that photoevaporation grows in strength to such a degree that it is sufficient to drive a monotonically \emph{decreasing} efficiency of star formation with decreasing metallicity.
Simulations by \cite{NakHos18}, however, show that the degree to which photoevaporation increases in strength with decreased metallicity, and indeed whether this strengthening of photoevaporation continues to arbitrarily low metallicity, is not a solved problem.

\section{Summary and Future Work}
\label{sec:summary}

In this work we have presented a parameter study of UV line-driven ablation of circumstellar discs around forming massive stars.
In the course of this study, we demonstrate that line-driven disc ablation proceeds at a rate well characterized by a constant enhancement factor over the stellar wind mass loss rate, namely $\dot{M}_\mathrm{ab} = 6.5\pm 1\; \dot{M}_\mathrm{wind}$.
Using this constant enhancement factor, we have predicted the disc mass loss rate as a function of stellar mass, as well as comparing this rate to characteristic accretion rates onto massive (proto)stars.
Of particular interest in this comparison is the inferred minimum accretion rates necessary to assemble stars of various masses.
Given the simple scaling of line-driven stellar wind mass loss rate with metallicity, we are also able to show the reduction in strength of ablation with decreasing metallicity, and the associated rise in the upper mass as predicted from only ablation.

As we have mentioned several times throughout this paper, however, this study and its conclusions on the nature of the stellar upper mass limit by design omit the impact of feedback mechanisms at larger scales, such as radiation pressure on dust grains, MHD driven outflows, and photoionization.
These feedback mechanisms operate on intrinsically larger scales than line-driven disc ablation, however, suggesting that the effects of ablation should add to the effects of feedback at larger scales by further reducing the accretion able to reach a massive (proto)star.
Additionally, this fundamental difference in scales should allow future studies of feedback at larger scales to incorporate line-driven ablation as a sub-grid model by adapting the results presented here.

As part of such an extension to larger scales, it would also be interesting to test what observables might be associated with line-driven ablation.
While the scales comparable to the stellar radius we have considered for the simulations in this work are fundamentally too small to be observed with present facilities, ablation is expected to set up a low latitude disc wind which could extend to observable scales.
Simulations using ablation as a sub-grid model could test the observability of this larger scale extension of the ablated disc wind, and compare to observations such as those published recently by Maud et al. (submitted), which argue for the presence of a disc wind traced in SiO emission, possibly consistent with such a layer.

In terms of future work that could improve the state of the art in our understanding of the mechanisms and effects at play in this study, a key step forward would be made by constraining the nature of accretion, specifically its latitudinal geometric distribution, for the high accretion rates predicted onto massive (proto)stars.
Such a study, even in the absence of ablation, would improve our understanding of how physics plays out in these final miles of accretion.
Moreover, the ability to self-consistently include both ablation and accretion in a single simulation would improve the constraints placed on the enhancement of ablation over wind mass loss, and directly test whether ablation depends on accretion rate or whether it is indeed as constant of an enhancement as the simulations presented here suggest.

Finally, future work should examine whether there is an additional dependence of ablation on stellar and or disc parameters beyond the simple scaling with the stellar wind mass loss rate.
Such a secondary, much weaker scaling is implied by comparing the $\dot{M}_\mathrm{ab} = 6.5 \pm 1 \; \dot{M}_\mathrm{wind}$ relation obtained here with the $\dot{M}_\mathrm{ab} = \dot{M}_\mathrm{wind}$ obtained for Classical Be stars in paper I.
Constraining whether this is due to the change in masses and spectral types considered from B stars to O stars, or if this is a by-product of the factor $\sim 1000$ increase in mass of discs from paper I to the results here would shed additional light on the nature of UV line-driven disc ablation.

\section*{Acknowledgements}

We would like to thank S. Owocki and J. Sundqvist for discussions related to the physics of line-driven stellar winds and their behavior at sub-solar metallicities.
We also would like to thank the anonymous referee for questions and comments clarifying the discussion in this paper.
This study was conducted within the Emmy Noether research group on ``Accretion Flows and Feedback in Realistic Models of Massive Star Formation" funded by the German Research Foundation under grant no. KU 2849/3-1.
Simulations were carried out using the computational resources available through the bwHPC initiative for high-performance computing in Baden-W\"{u}rttemberg, Germany.

\bibliographystyle{mn2e}
\bibliography{biblio}

\end{document}